\documentclass[twocolumn,showpacs,preprintnumbers,amsmath,amssymb,superscriptaddress,floatfix,nofootinbib]{revtex4}

\usepackage{graphicx}
\usepackage{amsmath}
\usepackage{amsfonts}
\usepackage{amssymb}

\begin{document}

\title{Study of $a_1(1260)$ in the $\gamma p \to \pi^+\pi^+\pi^- n$ reaction}
\date{\today}

\author{Xu Zhang}
\affiliation{Institute of Modern Physics, Chinese Academy of
Sciences, Lanzhou 730000, China} \affiliation{University of Chinese
Academy of Sciences, Beijing 101408, China}

\author{Ju-Jun Xie}~\email{xiejujun@impcas.ac.cn}
\affiliation{Institute of Modern Physics, Chinese Academy of
Sciences, Lanzhou 730000, China} \affiliation{University of Chinese
Academy of Sciences, Beijing 101408, China}

\begin{abstract}
We investigate the discovery potential of the $a_1(1260)$  photoproduction in the $\gamma p \to a_1(1260)^+ n$ and $\gamma p \to \pi^+\pi^+\pi^- n$ reactions
via the $\pi$-exchange mechanism.
For the $\gamma p \to \pi^+\pi^+\pi^- n$ reaction, we perform a calculation for the differential and total cross sections by including the contributions from the
$a_1(1260)$ intermediate resonance decaying into $\rho \pi$ then into $\pi^+\pi^+\pi^-$. Besides, the non-resonance process is also considered. With a lower mass of
$a_1(1260)$, we get a fairly good description of the experimental data for the invariant $\pi^+\pi^+\pi^-$ mass distributions. For the $\gamma p \to a_1(1260)^+ n$
reaction, with our model parameters, the total cross section is of the order of 10 $\mu$b at photon beam energy $E_{\gamma} \sim$ 2.5 GeV. It is expected that our model calculations could be tested by future experiments.
\end{abstract}

\maketitle

\section{INTRODUCTION}

The existence of a $J^{PC}=1^{++}$ meson is predicted by the quark model with SU(6)$\otimes$O(3) symmetry, which has made great success in classifying the hadrons ~\cite{Tanabashi:2018oca}. The $a_1(1260)$ with quantum numbers $J^{PC}=1^{++}$ is a candidate of the chiral partner of the $\rho$ meson ~\cite{Weinberg:1967kj,Bernard:1975cd,Ecker:1988te} described as a $q \overline{q}$ composite in the Numbu-Jona-Lasino model~\cite{Dhar:1983fr,Hosaka:1990sj} and in Lattice calculation~\cite{Wingate:1995hy}. Different from the quark model, it can be also considered as a gauge boson of the hidden local symmetry ~\cite{Bando:1987br,Kaiser:1990yf}, which is recently reconciled with the five-dimensional gauge field of the holographic QCD ~\cite{Sakai:2004cn,Sakai:2005yt,Nawa:2006gv}. On the other hand, by using the chiral unitary appraoch, $a_1(1260)$ is dynamically generated and can be interpreted as a quasibound state of pairs of hadrons in coupled channels~\cite{Roca:2005nm,Lutz:2003fm}.
The nature of $a_1(1260)$ has also been studied by calculating physical observables such as the $\tau$ decay spectrum into three pions ~\cite{GomezDumm:2003ku,Wagner:2008gz,Dumm:2009va} or multipions decays of light vector mesons~\cite{Achasov:2004re,Lichard:2006kw}.
Recently, the $a_1(1260)$ resonance is studied in Ref.~\cite{Dai:2018zki} in the decay of $\tau \to \nu_{\tau} \pi^- a_1(1260)$ through a triangle mechanism.

The dynamically generated nature of $a_1(1260)$ has been tested in the radiative decay process. The decay of $a_1(1260)$ into $\pi \gamma$ found in Ref.~\cite{Zielinski:1984au} were also studied in Ref.~\cite{Roca:2006am,Nagahiro:2008cv} and found to be in qualitative agreement with data if the $a_1(1260)$ is associated with the dynamically generated nature. In Ref.~\cite{Lang:2014tia} the lattice result for the coupling constant of the $a_1(1260)$ to the $\rho\pi$ channel is also close to the value obtained in Ref.~\cite{Roca:2005nm}.
Recently, the production of $a_1(1260)$ resonance in the reaction of $\pi^-p \to a_1(1260)^-p$ within an effective  Lagrangian approach  was studied in Ref.~\cite{Cheng:2016hxi} based on the results obtained in chiral unitary approach~\cite{Roca:2005nm}.
Furthermore, a general method was developed in Ref.~\cite{Nagahiro:2011jn} to analyze the mixing structure of hadrons consisting of two components of quark and hadronic composites, and the nature of the $a_1(1260)$ was explored with the method~\cite{Nagahiro:2011jn}, where it was found that the $a_1(1260)$ resonance has comparable amounts of the elementary component $q\bar{q}$ to the $\rho \pi$. In Ref.~\cite{Geng:2008ag}, the $N_c$ behavior of $a_1(1260)$ was studied using the unitarized chiral approach, and it was found that the main component of $a_1(1260)$ is not $q\bar{q}$. A probabilistic interpretation of the compositeness at the pole of a resonance has been derived in
Ref.~\cite{Guo:2015daa}, where it was obtained that, for $a_1(1260)$, the compositeness and elementariness are similar.

On the experimental side, for the $a_1(1260)$ resonance, the experimental Breit-Wigner width $\Gamma_{a_1(1260)}=(250-600)$ MeV assigned by
the Particle Data Group (PDG)~\cite{Tanabashi:2018oca} has large uncertainty. While most experiments and phenomenological extractions
agree on the mass of the $a_1(1260)$ leading to a PDG value of $M_{a_1(1260)}$ = 1230 $\pm$ 40 MeV, which is more precisely than
its width. A new COMPASS measurement in Ref.~\cite{Alekseev:2009aa} provides a much smaller uncertainty of the width $\Gamma_{a_1(1260)}
= 367 \pm 9 ^{+28} _{-25}$ MeV and mass $M_{a_1(1260)} = 1255 \pm 6^{+7}_{-17}$ MeV. Therefore, study of the photoproduction of $a_1(1260)$ is important both on experimental and theoretical sides, and can also provide beneficial information about
the internal structure of it.

Meson photoproduction off a baryon provides one of the most direct routes to extract information regarding the hadronic structure \cite{Huang:2016tcr,Xing:2018axn}.
We should point out that in the experiment, no signal representing $a_1(1260)^+ n$ photoproduction ~\cite{Condo:1993xa,Struczinski:1975ik,Ballam:1971wq,Nozar:2008aa,Eisenberg:1969kk} could be isolated even though the $\pi \gamma$ radiative width of $a_1(1260)$ very likely exceeds that of the $a_2(1320)$~\cite{Cihangir:1982ti,Zielinski:1984au,Molchanov:2001qk,May:1977ra}. The absence of this $J^{PC}=1^{++}$ state in charge exchange photoproduction is puzzled.
In this paper, by investigating the $\gamma p \to  a_1(1260)^+ n$ process within $\pi$-exchange mechanism, the total cross section is predicted. In our calculation, assuming that the $a_1(1260)$ resonance is dynamically generated state from pseudoscalar- meson-vector-meson interaction, the $\pi^+\pi^+\pi^-$ mass distribution and also the total cross section of $\gamma p \to \pi^+\pi^+\pi^- n$ are studied. In addition, we consider the non-resonance contributions to $\gamma p \to \pi^+\pi^+\pi^- n$ resonance which involve nucleon pole terms. Other  contributions which involve $\Delta(1232)$ and nucleon excited states can be removed based on the $\pi^+n$ invariants mass spectrum in experiment~\cite{Nozar:2008aa}.

The paper is organized as follows. After the introduction, we present the reaction mechanism of $a_1(1260)$ photoproduction. In Sec. \uppercase\expandafter{\romannumeral3}, the possible background relevant to the production of $a_1(1260)$ is discussed and the $\pi^+\pi^+\pi^-$ mass distribution is presented. This work ends with the discussion and conclusion.

\section{$\gamma p \to a_1(1260)^+ n$ reaction}

\begin{figure}[htbp]
\begin{center}
\includegraphics[scale=0.5]{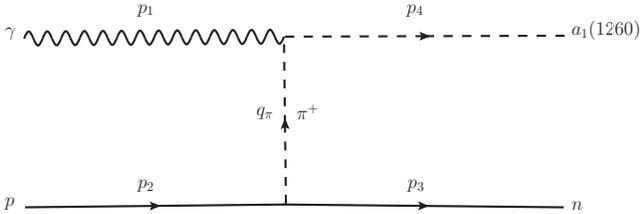}
\caption{Feynman diagram for the $\gamma p \to a_1(1260)^+ n$ reaction via $\pi$-exchange.} \label{fig:feynmana}
\end{center}
\end{figure}

In this section, we will discuss the production  mechanism of $a_1(1260)$ resonance.
Fig.~\ref{fig:feynmana} shows the basic tree-level Feynman diagram for the production of $a_1(1260)$ in a $\gamma p \to a_1(1260)^+ n$ reaction via $\pi$-exchange.

For the $\pi NN$ vertex we adopt the commonly used effective Lagrangian
\begin{eqnarray}
{\cal L}=&&-ig_{\pi NN}\bar{N}\gamma_5  (\vec{\tau} \cdot \vec{\pi} )N \nonumber\\
=&&-ig_{\pi NN}(\bar{p}\gamma_5p\pi^0+\sqrt{2}\bar{p}\gamma_5n\pi^+ \nonumber\\
&&+\sqrt{2}\bar{n}\gamma_5p\pi^--\bar{n}\gamma_5n\pi^0),
\end{eqnarray}
where a standard value, $g_{\pi NN}^2/{4\pi}=14.4$, is adopted as in Refs.~\cite{Machleidt:1987hj,Tsushima:1998jz}. In addition, a form factor is applied for the $\pi NN$ vertex,
\begin{eqnarray}
F_{\pi NN}(q_{\pi})=\frac{\Lambda_{\pi}^2-m_{\pi}^2}{\Lambda_{\pi}^2-q_{\pi}^2},
\end{eqnarray}
with $\Lambda_{\pi}$ the cut off parameter~\cite{Xie:2007vs,Xie:2007qt}, which will be discussed in the following.
$q_{\pi}$ is the momentum of exchanged $\pi$ meson.

The vertex depicting the interaction of $a_1(1260)$ and $\pi \gamma$ is~\cite{Roca:2006am,Nagahiro:2008cv}
\begin{equation}
t_{a_1^+\to \pi^+\gamma}= g_{a_1\pi\gamma} ( g^{\mu \nu}-\frac{p_{\gamma}^{\mu} p_{a_1}^{\nu}}{p_{\gamma} \cdot p_{a_1}} ) \varepsilon_{\mu}(p_{a_1})\varepsilon_{\nu}(p_{\gamma}),
\end{equation}
where $\varepsilon_{\mu}(p_{a_1})$ and $\varepsilon_{\nu}(p_{\gamma})$ are the polarization vector corresponding to $a_1(1260)$ and photon, respectively. Gauge invariance imposes a stringent constraint on radiative decay process of $a_1(1260)$ resonance. In the present case, a thorough test  of gauge invariance was conducted in Ref.~\cite{Nagahiro:2008cv} for the radiative decay of $a_1(1260)$ resonance within the local hidden gauge approach.

The coupling constant $g_{a_1\pi\gamma}$ can be determined by the decay width of $\Gamma_{a_1\to \pi \gamma}$. With the vertex above, we obtain
\begin{eqnarray}
\Gamma_{a_1\to \pi \gamma}=\frac{g_{a_1\pi\gamma}^2}{24\pi M_{a_1}^3}(M_{a_1}^2-m_{\pi}^2),
\end{eqnarray}
where $M_{a_1}=1230$ MeV is the nominal mass of the $a_1(1260)$.
Using the partial decay width  $\Gamma_{a_1\to \pi \gamma}=640\pm246$ keV of $a_1(1260)$ as listed in the PDG book~\cite{Tanabashi:2018oca}, we get $g_{a_1\pi\gamma}=244 \pm 94$ MeV, where the error is from the uncertainties of $\Gamma_{a_1\to \pi \gamma}$ and the mass of $a_1(1260)$. In the following calculations, we take the average value $g_{a_1\pi\gamma}=244$ MeV.

With the above integrants, we finally obtain the amplitude of $\gamma(p_1) p(p_2) \to a_1(1260)^+(p_4)+ n(p_3)$ process as
\begin{eqnarray}
{\cal M}=&&\frac{-\sqrt{2}ig_{\pi NN}g_{a_1 \pi \gamma}}{q_{\pi}^2-m_{\pi}^2}\bar{u}(p_3)\gamma_5u(p_2)\times \nonumber\\
&&( g^{\mu \nu}-\frac{p_1^{\mu} p_4^{\nu}}{p_1 \cdot p_4} ) \varepsilon_{\mu}(p_4)\varepsilon_{\nu}(p_1) F_{\pi NN}(q_{\pi}).
\end{eqnarray}
By defining $s=(p_1+p_2)^2$, the corresponding unpolarized differential cross section reads as
\begin{eqnarray}
\frac{d\sigma}{d cos\theta}=\frac{1}{32\pi s}\frac{\vert \vec{p}_4^{\ c.m.}\vert}{\vert \vec{p}_1^{\ c.m.}\vert}\left(\frac{1}{4}\sum_{spins}\vert {\cal M}\vert^2 \right),
\end{eqnarray}
where $\theta$ denotes the angle of outgoing $a_1^+$ meson relative to the beam direction in the c.m. frame, while $\vec{p}_1^{\ c.m.}$ and $\vec{p}_4^{\ c.m.}$ are the three-momenta of the initial photon beam and  the final $a_1^+$, respectively.

\begin{figure}[htbp]
\begin{center}
\includegraphics[scale=0.9]{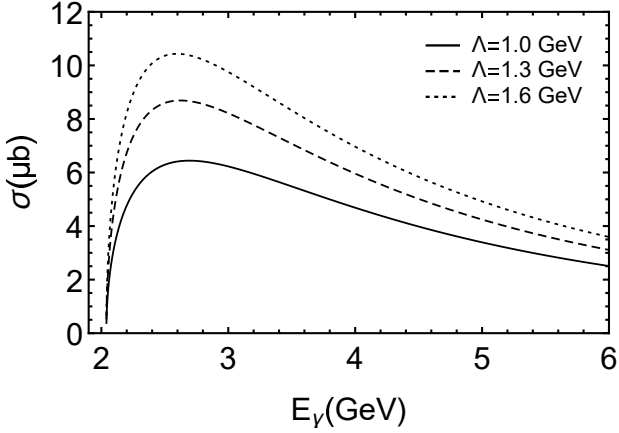}% Here is how to import EPS art
\caption{The dependence of total cross section of $\gamma p \to a_1(1260)^+ n$ as a function of $E_{\gamma}$.} \label{fig:section}
\end{center}
\end{figure}

In  Fig.~\ref{fig:section}, the solid, dashed and dotted lines are obtained with $\Lambda_{\pi}=1.0$, 1.3 and 1.6 GeV, respectively.
From Fig.~\ref{fig:section} it is seen that the total cross section via $\pi$ exchange goes up very rapidly near the threshold, and the peak position of the total cross section is $E_{\gamma} = 2.6$ GeV. The total cross section is proportional to $g_{a_1\pi\gamma}^2$, which indicates that the cross section is proportional to the partial decay width of $\Gamma_{a_1\to \pi \gamma}$. Since the concrete value of $\Gamma_{a_1\to \pi \gamma}$ is undetermined by theory and experiment, in this work we take $\Gamma_{a_1\to \pi \gamma}=640$ keV. The result is comparable with the cross section of $a_2(1320)$ photoproduction~\cite{Huang:2013jda}.

\section{$\gamma p \to \pi^+\pi^+\pi^- n$ reaction}

Next, we pay attention to
the $\gamma p\to a_1(1260)^+n\to\rho^0\pi^+n\to\pi^+\pi^+\pi^-n$ and $\gamma p\to \rho^0p\to\pi^+\pi^+\pi^-n$ processes. Here $\gamma p\to \rho^0p\to\pi^+\pi^+\pi^-n$ can occur via nucleon pole term~\cite{Huang:2014gxa}.

\subsection{$\gamma p\to a_1(1260)^+n\to\rho^0\pi^+n\to\pi^+\pi^+\pi^-n$ REACTION}

\begin{figure}[htbp]
\begin{center}
\includegraphics[scale=0.55]{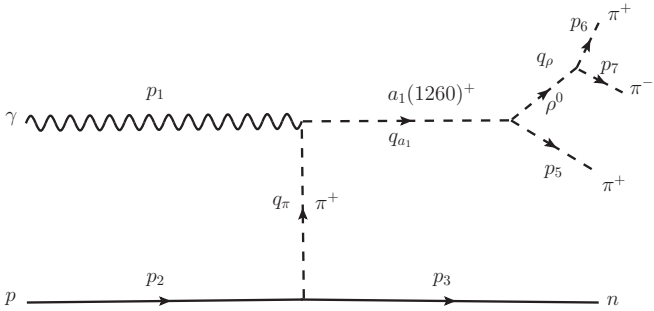}
\caption{Feynman diagram for the $\gamma p\to a_1(1260)^+n\to\rho^0\pi^+n\to\pi^+\pi^+\pi^-n$ reaction via $\pi$ exchange.} \label{fig:feynmanb}
\end{center}
\end{figure}

The $\gamma p\to a_1(1260)^+n\to\rho^0\pi^+n\to\pi^+\pi^+\pi^-n$ reaction with the $\pi$ exchange is described in Fig.~\ref{fig:feynmanb}, where the relevant kinematic variables are marked. As mentioned in the introduction, we will consider the $a_1(1260)$ as a dynamically generated state in coupled channels of $\rho \pi$ and $\bar{K}^*K$.

The $a_1^+\rho^0\pi^+$ vertex can be written as
\begin{eqnarray}
-it_1=-i\frac{g_{a_1\rho\pi}}{\sqrt{2}}\varepsilon_{a_1}^{\mu} \varepsilon_{\mu}
\end{eqnarray}
where $\varepsilon_{a_1}$ is the polarization vector of $a_1(1260)$ and  $\varepsilon$ the polarization vector of the $\rho$. The $g_{a_1\rho\pi}$ is the coupling of the $a_1(1260)$ to the $\rho\pi$ and can be obtained from the residue in the pole of the scattering amplitude in $I=1$. We take $g_{a_1\rho\pi}=(-3795+i2330)$ MeV as obtained in Ref.~\cite{Roca:2005nm}.

For the vertex of $a_1(1260)^+$ interacting with $\rho^0\pi^+$, we also introduce a form factor $F_{a_1\rho\pi}$, which is
\begin{eqnarray}
F_{a_1\rho\pi}(q_{a_1})=\frac{\Lambda_{a_1}^4}{\Lambda_{a_1}^4+(q_{a_1}^2-M_{a_1}^2)^2},
\end{eqnarray}
with a typical value of $\Lambda_{a_1}=1.5$ GeV as used in Refs.~\cite{Cheng:2016hxi,Zhang:2018tko}.

The $a_1(1260)$  propagator is
\begin{eqnarray}
G_{a_1}^{\alpha \beta}(q_{a_1})=i \frac{-g^{\alpha \beta}+q_{a_1}^{\alpha}q_{a_1}^{\beta}/M_{a_1}^2}{q_{a_1}^2-M_{a_1}^2+iM_{a_1}\Gamma_{a_1}},
\end{eqnarray}
where the width $\Gamma_{a_1}$  is dependent on its four-momentum squared, and we can take the form as in Refs.~\cite{Xie:2014twa,Xie:2015isa},
\begin{eqnarray}
\Gamma_{a_1}=\Gamma_0+\Gamma_{3\pi},
\end{eqnarray}
where $\Gamma_{3\pi}$ is the decay width for the process $a_1(1260)\to \rho\pi \to 3\pi$ \cite{Zhang:2018tko}, and $\Gamma_0$ is the decay width for other processes. Following the experiment result in Ref.~\cite{Alekseev:2009aa} for the total decay width of $a_1(1260)$, we take $\Gamma_{0}=201$ MeV for $\Gamma_{a_1}=367$ MeV at $\sqrt{q_{a_1}^2}=1230$ MeV.

The structure of the vertex for the $PPV$ interaction can be evaluated by means of hidden gauge symmetry Lagrangian \cite{Bando:1984ej,Bando:1987br,Harada:2003jx}
\begin{eqnarray}
{{\cal L}_{PPV}}=-ig<V^{\mu}[P,\partial_{\mu}P]>, \label{eq:hidden}
\end{eqnarray}
where the $<>$ stands for the trace in $SU(3)$ and $g=\frac{m_{V}}{2f}$, with $m_{V}=m_{\rho}$ and $f=93$ MeV the pion decay constant. The matrices $P$ and $V$ in Eq.~(\ref{eq:hidden}) contain the nonet of the pseudoscalar mesons and the one of the vectors respectively. The resulting amplitude for the vertex can be written as
\begin{eqnarray}
{-it}=-i\sqrt{2}g(p_7-p_6)_{\lambda}\varepsilon^{\lambda}(p_4),
\end{eqnarray}
For the vertex of $\rho$ interacting with $\pi\pi$, we also introduce a form factor $F_{\rho\pi\pi}$, which satisfies the form
\begin{eqnarray}
F_{\rho\pi\pi}(q_{\rho})=\frac{\Lambda_{\rho}^4}{\Lambda_{\rho}^4+(q_{\rho}^2-m_{\rho}^2)^2},
\end{eqnarray}
with a typical value of $\Lambda_{\rho}=1.5$ GeV as used in Ref.~\cite{Zhang:2018tko}.

The $\rho$ propagator is
\begin{eqnarray}
G_{\rho}^{\sigma\lambda}(q_{\rho})=i \frac{-g^{\sigma \lambda}+q_{\rho}^{\sigma}q_{\rho}^{\lambda}/m_{\rho}^2}{q_{\rho}^2-m_{\rho}^2+im_{\rho}\Gamma_{\rho}},
\end{eqnarray}
with the energy dependent decay width of $\Gamma_\rho$. Because the dominant decay channel of $\rho$ is $\pi \pi$, we take
\begin{eqnarray}
\Gamma_{\rho}(M^2_{\rm inv})=\Gamma_{{\rm on}}\left (\frac{q_{{\rm
off}}}{q_{{\rm on}}}\right )^3 \frac{m_{\rho}}{M_{\rm inv}},
\label{eq:gamrrhopipi}
\end{eqnarray}
with $\Gamma_{\rm on} = 149.1$ MeV, and
\begin{eqnarray}
q_{\rm on} &=& \frac{\sqrt{m^2_\rho - 4m^2_\pi}}{2}, \\
q_{\rm off} &= & \frac{\sqrt{M^2_{\rm inv} - 4m^2_\pi}}{2},
\end{eqnarray}
with $M^2_{\rm inv}=q_{\rho}^2=(p_6+p7)^2$ or $(p_5+p_7)^2$ the invariant mass square of the $\pi^+ \pi^-$ system. We take $m_\rho = 775.26$ MeV in this work.

It is worthy to mention that the parametrization of the width of the $\rho$ meson shown in Fig.~\ref{fig:feynmanb} is common and it is
meant to take into account the phase space of each decay mode as a function of the
energy~\cite{Chiang:1990ft,Xie:2007qt,Hanhart:2010wh}. In the present work we take explicitly the phase space for the $P$-wave
decay of the $\rho$ into two pions.

With the above preparation, we finally obtain the scattering amplitude for the diagram shown in Fig.~\ref{fig:feynmanb},
\begin{eqnarray}
{\cal M}_{\uppercase\expandafter{\romannumeral1}}=&&
\frac{\sqrt{2}ig_{\pi NN}g_{a_1 \pi \gamma}}{q_{\pi}^2-m_{\pi}^2}\bar{u}(p_3)\gamma_5u(p_2)( g^{\mu \nu}-\frac{p_1^{\mu} q_{a_1}^{\nu}}{p_1 \cdot q_{a_1}} )  \nonumber\\
&&\varepsilon_{\nu}(p_1)G^{a_1}_{\mu\sigma}(q_{a_1}) F_{\pi NN}(q_{\pi})F_{a_1\rho\pi}(q_{a_1})(g_{\rho\pi}g) \nonumber\\
 &&\Big( G_{\rho}^{\sigma\lambda}(p_6+p_7)(p_7-p_6)_{\lambda}F_{\rho\pi\pi}(p_6+p_7)+\nonumber\\
 &&( G_{\rho}^{\sigma\lambda}(p_5+p_7)(p_7-p_5)_{\lambda}F_{\rho\pi\pi}(p_5+p_7) \Big).
\end{eqnarray}

\subsection{$\gamma p\to \rho^0p\to\pi^-\pi^+p\to\pi^+\pi^+\pi^-n$ REACTION}

\begin{figure}[htbp]
\begin{center}
\includegraphics[scale=0.55]{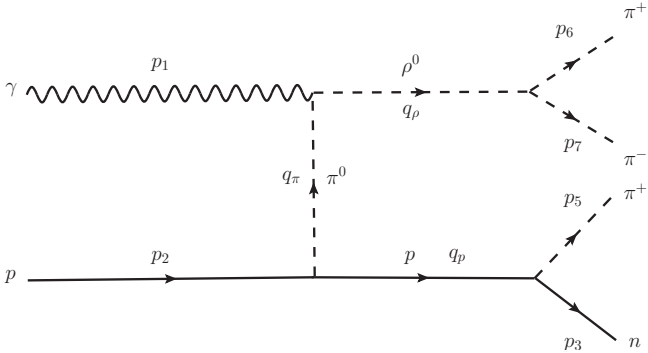}% Here is how to import EPS art
\caption{Feynman diagram for the $\gamma p \to\rho^0\pi^+n\to\pi^+\pi^+\pi^-n$ reaction via $\pi$ exchange.} \label{fig:feynmanc}
\end{center}
\end{figure}

Besides the resonance contribution from $a_1(1260)$ resonance, we study another kind of reaction mechanism for $\gamma p \to\pi^+\pi^+\pi^-n$ reaction, which is depicted in Fig.~\ref{fig:feynmanc},
where we have considered the contribution from $\gamma p \to\rho^0 p \to\pi^+\pi^-\pi^+n$. In Fig.~\ref{fig:feynmanc}, the relevant kinematic variables are also shown.

To compute the contribution of Fig.~\ref{fig:feynmanc}, we take the interaction density for $\rho \gamma \pi$ as~\cite{Oh:2003aw,Oh:2000zi},
\begin{eqnarray}
{\cal L_{\rho\gamma\pi}}=\frac{eg_{\rho\gamma\pi}}{m_{\rho}}\epsilon^{\mu\nu\alpha\beta}\partial_{\mu}\rho_{\nu}\partial_{\alpha}A_{\beta}\pi,
\end{eqnarray}
where $A_{\beta},\pi$ and $\rho_{\nu}$ denote the fields for the photon, $\pi$ and $\rho$,
respectively. The coupling constant of $g_{\rho\gamma\pi}$ can be obtained by the experiment decay width
of $\Gamma_{\rho^0 \to \pi^0\gamma}$~\cite{Tanabashi:2018oca} which leads to $g_{\rho\gamma\pi}=0.76$.

Other vertexes are same as given above. With the above preparation, we finally obtain the transition amplitude for the diagram shown in Fig.~\ref{fig:feynmanc},
\begin{eqnarray}
&&{\cal M}_{\uppercase\expandafter{\romannumeral2}}=
\frac{-\sqrt{2}g_{\pi NN}g_{\pi NN}}{q_{\pi}^2-m_{\pi}^2}\frac{eg_{\rho\gamma\pi}}{m_{\rho}}gF_{\pi NN}(q_{\pi})\bar{u}(p_3)\gamma_5 \nonumber\\
&&\Big(\frac{({p_3}\!\!\!\!\!/+{p_5}\!\!\!\!\!/ \ )+m_p}{(p_3+p_5)^2-m_p^2}\gamma_5 u(p_2)F_{\pi NN}(p_3+p_5)\epsilon^{\mu\nu\alpha\beta}\nonumber\\
&&(p_6+p_7)_{\alpha}p_{1\beta}\epsilon_{\nu} G_{\rho}^{\mu\sigma}(p_6+p_7)(p_7-p_6)_{\sigma}F_{\rho\pi\pi}(p_6+p_7)\nonumber\\
&&+\frac{({p_3}\!\!\!\!\!/+{p_6}\!\!\!\!\!/ \ )+m_p}{(p_3+p_6)^2-m_p^2}\gamma_5 u(p_2)F_{N}(p_3+p_6)\epsilon^{\mu\nu\alpha\beta}(p_5+p_7)_{\alpha}\nonumber\\
&&p_{1\beta}\epsilon_{\nu} G_{\rho}^{\mu\sigma}(p_5+p_7)(p_7-p_5)_{\sigma}F_{\rho\pi\pi}(p_5+p_7)\Big),
\end{eqnarray}
with
\begin{eqnarray}
F_{N}(q_{p})=\frac{\Lambda_{N}^4}{\Lambda_{N}^4+(q_{p}^2-m_{p}^2)^2},
\end{eqnarray}
where $\Lambda_{\pi}=0.6$ GeV and $\Lambda_{N}=0.5$ GeV are taken from Refs.~\cite{Oh:2003aw,Oh:2000zi,Titov:1999eu}.
This choice of the cut off leads to a satisfactory explanation of the $\rho^0$ photoproduction at low energies.
Note that the value of $\Lambda_{\pi}$ is different with the one we used before for the $\gamma p \to na_1(1260)^+$ production.
Other cut off parameters are same as given above.

\begin{figure*}[htbp]
\begin{center}
  \includegraphics[scale=0.9]{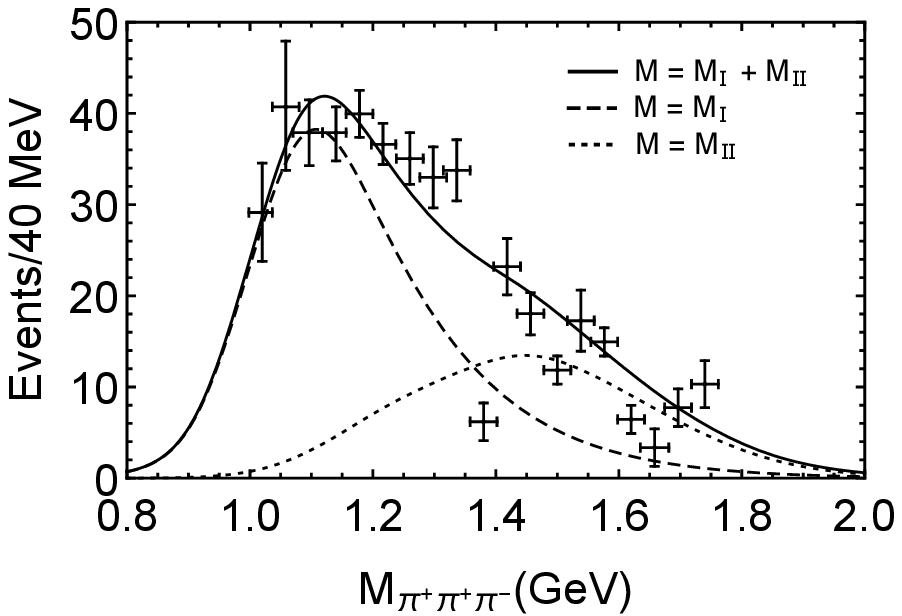}
\includegraphics[scale=0.9]{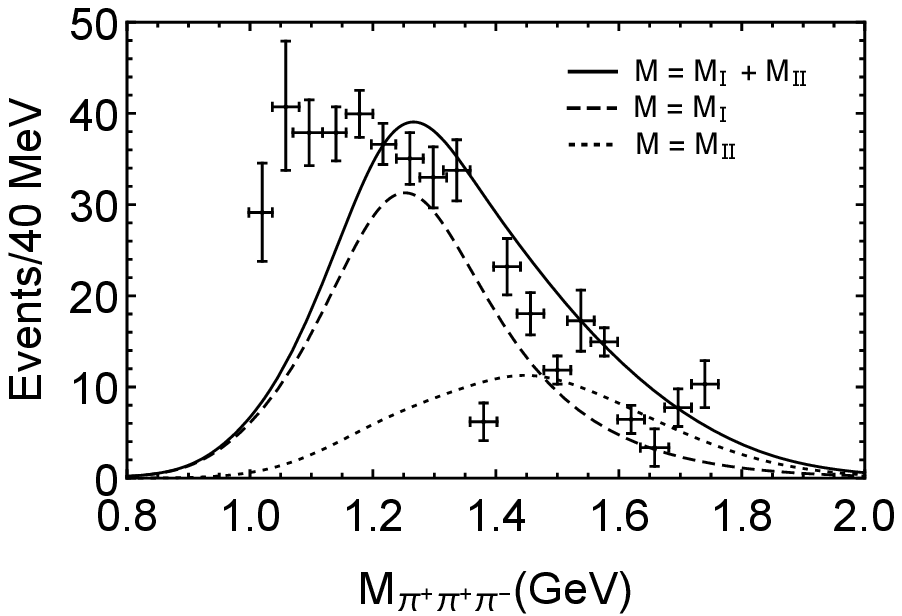}
\caption{The 3 $\pi$ invariant mass spectrums for the $\gamma p\to \pi^+\pi^+\pi^-n$ process are compared with the date obtained by the CLAS Collaboration from $1^{++}(\rho\pi)_S$ partial wave \cite{Nozar:2008aa}. Left and right plot correspond to  $M_{a_1}=1080$ and $1230$ MeV respectively.} \label{fig:spectrums}
\end{center}
\end{figure*}

\begin{figure*}[htbp]
\begin{center}
  \includegraphics[scale=0.9]{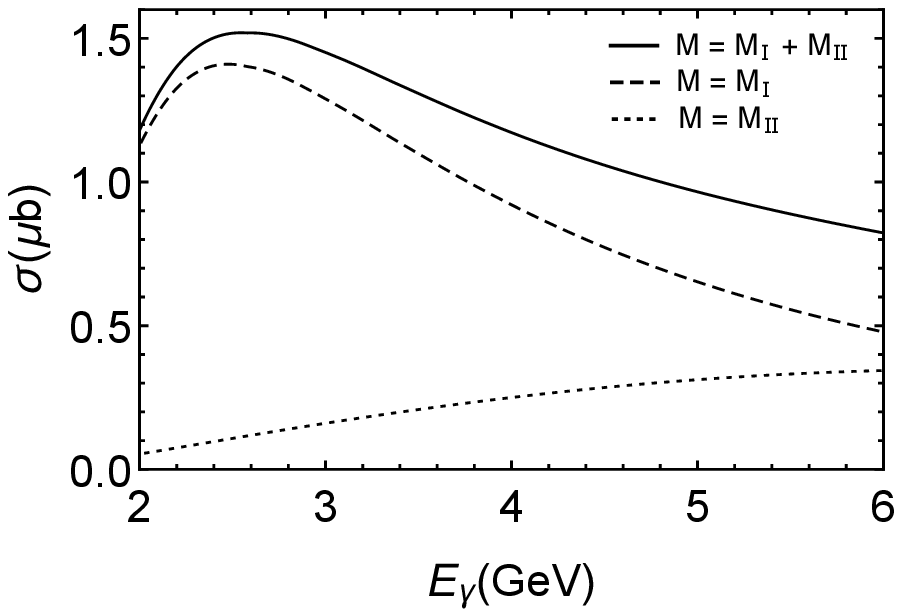}
  \includegraphics[scale=0.9]{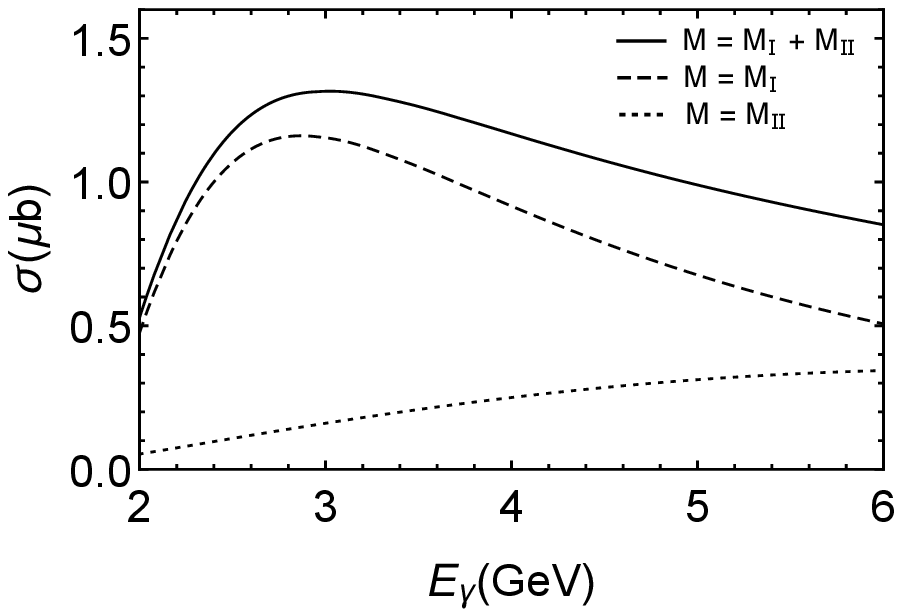}
\caption{Total cross section for the $\gamma p\to \pi^+\pi^+\pi^-n$ process are compared with the date obtained by the CLAS Collaboration from $1^{++}(\rho\pi)_S$ partial wave \cite{Nozar:2008aa}. Left and right plot correspond to  $M_{a_1}=1080$ and $1230$ MeV respectively.} \label{fig:sigmav}
\end{center}
\end{figure*}

\subsection{NUMERICAL RESULTS}
Then, the total cross section of the  $\gamma p\to\pi^+\pi^+\pi^-n$ reaction could be obtained by integrating the invariant amplitude in four-body phase space:
\begin{eqnarray}
&d\sigma(\gamma p \to \pi^+\pi^+\pi^-n )=\frac{1}{2!}\frac{2m_p\cdot2 m_n}{4\vert p_1\cdot  p_2 \vert}\left(\frac{1}{4}\sum\limits_{spins}\vert {\cal M}\vert^2 \right)\nonumber\\
&\times (2\pi)^4d\phi_4(p_1+p_2;p_3;p_5,p_6,p_7),
\end{eqnarray}
with
\begin{eqnarray}
{\cal M}={\cal M}_{\uppercase\expandafter{\romannumeral1}}+{\cal M}_{\uppercase\expandafter{\romannumeral2}},
\end{eqnarray}
where 2! is statistical factor for the final two $\pi^+$ mesons,
and the four-body phase space is defined as \cite{Tanabashi:2018oca}
\begin{eqnarray}
&d\phi_4(p_1+p_2;p_3;p_5,p_6,p_7)=- \frac{1}{16(2\pi)^8\sqrt{s}}
 \vert \vec{p}_6^{\ *a}\vert \vert \vec{p}_5^{\ *b} \vert  \vert \vec{p_3} \vert\nonumber\\
&d\Omega_6^{*a}  d\Omega_5^{*b} d\Omega_3  dM_{\pi^+\pi^-} dM_{\pi^+\pi^+\pi^-},
\end{eqnarray}
where $\vert\vec{p}_6^{\ *a}\vert$ and $\Omega_6^{*a}$ are the three-momentum and solid angle of the out going $\pi^+$ in c.m. frame of the final $\pi^+\pi^-$ system, and $ \vert \vec{p}_5^{\ *b} \vert$ and $\Omega_5^{*b}$ are the three-momentum and solid angle of the out going $\pi^+$ in c.m. frame of the final $\pi^+\pi^+\pi^-$ system, and $\vert \vec{p}_3 \vert$ and $\Omega_3$ are the three-momentum and solid angle of the out going $n$ in c.m. frame of the initial $\gamma p$ system. In the above equation, $M_{\pi^+\pi^-}$ is the invariant mass of the $\pi^+\pi^-$ two body system, and $M_{\pi^+\pi^+\pi^-}$ is the invariant mass of the $\pi^+\pi^+\pi^-$ three body system, and $s=(p_1+p_2)^2$ is the invariant mass square of the initial $\gamma p$ system.

In Ref.~\cite{Nozar:2008aa}, the $\gamma p\to\pi^+\pi^+\pi^-n$ reaction was studied in the photon energy range 4.8-5.4 GeV.  The 3$\pi$ mass distributions are measured from $1^{++}(\rho\pi)_S$ partial wave.
In Fig.~\ref{fig:spectrums}, we show the theoretical results, $c_1d\sigma/dM_{\pi^+\pi^+\pi^-}$, for the $\pi^+\pi^+\pi^-$ invariant mass distributions for the $\gamma p\to\pi^+\pi^+\pi^-n$ reaction at $E_{\gamma}=5.1$ GeV, compared with the experimental measurements of Ref.~\cite{Nozar:2008aa}. The theoretical
results are obtained with $c_1=21.5$ and $c_1=18$ for $M_{a_1}=1080$ and $1230$ MeV, respectively, which have been adjusted to the strength of the experimental data reported by the CLAS collaboration \cite{Nozar:2008aa}.
From Fig.~\ref{fig:spectrums}, it is seen that the bump structure around $1.4-1.6$ GeV may account for the nuclear pole contribution.
If we use a mass $M_{a_1}=1080$ MeV, $\pi^+\pi^+\pi^-$ invariant mass distributions agree with the experimental data well.
On the other hand, the theoretical results with $M_{a_1}=1230$ MeV can not describe the bump structure around 1.1 GeV.

In addition to the differential cross section, we calculate also the total cross section for the $\gamma p\to \pi^+\pi^+\pi^-n$ process as a function of the photon beam energy $E_{\gamma}$. The result are shown in Fig.~\ref{fig:sigmav},
where one can see that the total cross section goes up rapidly near the threshold, and the peak position of the total cross section is $E_{\gamma}=2.5$ and 2.9 GeV corresponding to $M_{a_1}=1080$ and $1230$ MeV, respectively. The differential and total cross section can be checked in future experiments, such as those at the CLAS.

\section{CONCLUSION AND DISCUSSION}

In recent years, it has been found that the $a_1(1260)$ resonance, although long accepted as an ordinary $q\overline{q}$ state, can be dynamically generated from pseudoscalar- meson-vector-meson interaction, and therefore qualify as a pseudoscalar-vector molecule. In this work, we have proposed to test the molecular picture in the photoproduction process. Since $a_1(1260)$ was observed in the radiative decay of $a_1(1260)^+\to \pi^+\gamma$, the $\gamma p \to a_1(1260)^+ n$ reaction by exchanging the $\pi$ meson is the main process to produce $a_1(1260)$. Our numerical results show that the total cross section of $\gamma p \to a_1(1260)^+ n$ is of a order of 10 $\mu$b, which is comparable with the cross section of the photoproduction of $a_2(1320)$.

Additionally, assuming that the $a_1(1260)$ resonance is a dynamically generated state from pseudoscalar- meson-vector-meson interaction, the $\pi^+\pi^+\pi^-$ mass distributions of $\gamma p \to \pi^+\pi^+\pi^- n$ are studied. With $M_{a_1}=1080$ MeV, we can describe the experimental data on the invariant $\pi^+\pi^+\pi^-$ invariant mass distributions fairly well. The total cross sections of $\gamma p \to \pi^+\pi^+\pi^- n$ reaction are also studied with these model parameters determined from the comparing with the experimental  data on the $\pi^+\pi^+\pi^-$ invariant mass distributions. It is expected that our model calculations could be tested by future experiments about
$\gamma p\to \pi^+\pi^+\pi^-n$ reaction at the photon beam energy $E_{\gamma}$ around 2.5 $\sim$ 2.9 GeV.

\section*{Acknowledgments}

One of us (Xu Zhang) would like to thank Yin Huang for helpful discussions. This work is partly supported by the National Natural Science Foundation of China under Grant Nos. 11475227 and 11735003, and by the Youth Innovation Promotion Association CAS (No. 2016367).

\end{document}